# The Initial Mass Function of the Galactic Early-type Field Stars Based on the LAMOST Survey

Qida Li[1,2], Jianping Xiong[1], Zhenwei Li[1], Dan Qiu[2,3], Chao Liu[2,3], Zhanwen Han[1], and Xuefei Chen[1,4]
[1] Yunnan Observatories, Chinese Academy of Sciences, Kunming, 650216, People's Republic of China; cxf@ynao.ac.cn
[2] School of Astronomy and Space Science, University of Chinese Academy of Sciences, Beijing, 100049, People's Republic of China
[3] Key Laboratory of Space Astronomy and Technology, National Astronomical Observatories, Chinese Academy of Sciences, Beijing 100101, People's Republic of China
[4] International Centre of Supernovae, Yunnan Key Laboratory, Kunming, 650216, People's Republic of China

## Abstract

Research on the high-mass end of the initial mass function (IMF) has been limited due to a scarcity of samples. Recently, Large Sky Area Multi-Object Fiber Spectroscopic Telescope (LAMOST), as the most efficient spectroscopic telescope, has provided new opportunities for related research. In this study, based on approximately 70,000 main-sequence early-type stars from the LAMOST survey, we investigated the IMF of Galactic field stars at the high-mass end ($1.5 \leqslant M/M_\odot \leqslant 7.1$). First, we derived the slope of the present-day mass function (PDMF), finding $\alpha_{\rm pre} = 5.59 \pm 0.02$ after correcting for selection effect in the observed sample. We then accounted for the effects of stellar evolution and unresolved binaries to correct the PDMF back to the IMF, resulting in $\alpha_{\rm ini} = 2.70 \sim 2.82$. Notably, we corrected both stellar evolution and unresolved binary effects simultaneously by using binary-Star evolution code, which enhances the robustness of our results. Additionally, we investigated how different mass-ratio ($q$) distributions of binaries and different star formation histories of the Milky Way impact the IMF. Finally, we tested samples across different spatial scales and found that $\alpha_{\rm ini}$ may exhibit a decreasing trend as the spatial scale increases, which could be attributed to variations in metallicity.

*Unified Astronomy Thesaurus concepts:* Initial mass function (796); Early-type stars (430); Stellar evolution (1599)

## 1. Introduction

The initial mass function (IMF) describes the mass distribution during star formation events. Initial mass and metallicity are critical parameters in stellar evolution, as they determine the evolutionary path and ultimate fate of a star. The mass distribution within a star cluster sets its kinematics, so the evolution of a coeval star cluster is primarily affected by its stellar IMF. Similarly, the total luminosities and chemical evolution of galaxies depend on the stellar IMF and star formation history (SFH; A. M. Hopkins 2018). The concept of the IMF was first proposed by E. E. Salpeter (1955), who discovered that the mass distribution of stars in the solar neighborhood follows a power law in the mass range of $0.4 \leqslant M/M_\odot \leqslant 10$, with a power-law exponent of $\alpha = 2.35$. This suggests that low-mass stars are more numerous than high-mass stars. As research progressed, it became evident that the slope is flatter, with $\alpha \approx 1.3$, for low-mass stars and steeper, with $\alpha \approx 2.7$, for high-mass stars (G. E. Miller & J. M. Scalo 1979; S. Basu & N. C. Rana 1992; P. Kroupa et al. 1993; I. N. Reid et al. 2002; K.-P. Schröder & B. E. J. Pagel 2003; N. Bastian et al. 2010; P. Kroupa et al. 2013). This indicates the presence of a breakpoint in the IMF, which is generally considered to occur around $1\,M_\odot$ (A. M. Hopkins 2018). Furthermore, P. Kroupa & C. Weidner (2003) found that the IMF of high-mass field stars is steeper than that of stars in clusters.

Clusters are ideal laboratories for studying the IMF, as stars within the same cluster share similar chemical abundances and ages. Therefore, SFH does not need to be assumed. For young massive clusters, the effects of stellar evolution on the IMF are significantly less pronounced compared to field stars. Several previous studies (D. F. Figer et al. 1999; H. Sung & M. S. Bessell 2004; B. Lim et al. 2013; M. W. Hosek et al. 2019) investigated young massive star clusters (age <5 Myr) and found that $\alpha \approx 1.8$. However, the influence of binary systems was not considered in these studies. Furthermore, P. Kroupa & C. Weidner (2003) showed that the field-star IMF tends to be steeper than the stellar IMF of local star-forming regions such as clusters for stars with $M/M_\odot > 1$. It is considered that massive clusters, which produce most of the massive stars, are rare. Consequently, their contribution to the field-star IMF is limited, leading to a steeper slope for the field-star IMF compared to that of individual clusters. The IMF of field stars includes stars born at different epochs and spans a broader range of metallicities, providing a more comprehensive picture of the mass distribution in the Milky Way. Studying the IMF of field stars enables us to predict the abundance of neutron stars and black holes and to better understand the chemical evolution of the Milky Way.

For the IMF of field stars, extensive research has been conducted on the low-mass end (D. S. Graff & K. Freese 1996; I. N. Reid & J. E. Gizis 1997; R. D. Jeffries 2012; K. L. Luhman 2012; J. Li et al. 2023). J. Li et al. (2023) demonstrate that the IMF slope depends on metallicity and stellar age. In contrast, research on the high-mass end of the IMF is relatively limited. J. M. Scalo (1986) summarized early work on the IMF of high-mass stars. However, due to small sample sizes and large measurement uncertainties, they found $\alpha = 2.3 \sim 3.4$ for stars with masses greater than $15\,M_\odot$. I. N. Reid et al. (2002) extended the analysis by using 558 main-sequence stars and assuming a constant SFH, resulting in a high-mass end slope of $\alpha = 2.5 \sim 2.8$. K.-P. Schröder & B. E. J. Pagel (2003) also studied the massive field stars of the Milky Way by considering

SFH and found $\alpha = 2.7 \pm 0.15$ for $1.1 \leqslant M/M_\odot \leqslant 1.6$, and $\alpha = 3.1 \pm 0.15$ for $1.6 \leqslant M/M_\odot \leqslant 4$. P. Kroupa et al. (2013) presented the Galactic-field stellar IMF, reporting $\alpha = 2.7 \pm 0.4$ for stars with masses greater than $1 M_\odot$. J. B. Lamb et al. (2013) studied the IMF of field stars using 284 OB-type stars ($M > 20 M_\odot$) in the Small Magellanic Cloud, finding $\alpha = 3.3 \pm 0.4$. However, a common limitation of these studies is the small sample size.

Moreover, many of the observed massive stars have already evolved off of the main sequence, so correction for evolutionary effects is needed (I. N. Reid et al. 2002; K.-P. Schröder & B. E. J. Pagel 2003; B. G. Elmegreen & J. Scalo 2006). The inclusion of an SFH is typically required, which introduces further uncertainties in the IMF measurement. In addition to this, the study of the IMF requires accounting for the influence of unresolved binary systems (R. Sagar & T. Richtler 1991; C. Weidner et al. 2009; J. J. Bochanski et al. 2010; K. L. Luhman 2012; O. De Marco & R. G. Izzard 2017). The impact of unresolved binaries is twofold: (1) the presence of a companion star contributes additional luminosity, resulting in erroneous mass measurement; and (2) star counts are affected, as both binary members should be included in the statistical analysis. Given the high binary fraction characteristic of early-type stars (H. Sana et al. 2012; Y. Guo et al. 2022a, 2022b), the binary evolution differs significantly from the evolutionary paths of single stars. Therefore, considering the evolutionary processes of binary systems is crucial in studies of the IMF.

The rise of large-scale surveys like the Large Sky Area Multi-Object Fiber Spectroscopic Telescope (LAMOST; X.-Q. Cui et al. 2012; L.-C. Deng et al. 2012; X.-W. Liu et al. 2014), providing a wealth of early-type star data, has presented a significant opportunity for advancements in related research. For instance, Y. Guo et al. (2021) and M. Xiang et al. (2022) used machine learning to measure atmospheric parameters for early-type stars in the LAMOST low-resolution spectral data, providing catalogs of 16,032 and 332,172 stars, respectively. Based on their derived atmospheric parameters, Q. D. Li et al. (2025) performed mass predictions for 3657 and 132,624 main-sequence early-type stars (S/N > 15), respectively, with an uncertainty of approximately 9%. Furthermore, Gaia provides a complete all-sky data set that can be used to correct for the completeness of the LAMOST sample. These abundant data provide rich and reliable resources for exploring the IMF of early-type stars.

Therefore, this paper aims to investigate the high-mass end of the IMF using early-type stars from the LAMOST survey, taking into account the evolutionary effect and unresolved binary effect. The structure of this paper is as follows: Section 2 presents the early-type star sample employed in this study. Section 3 corrects the selection effect for the observational sample and derives the present-day mass function (PDMF). Section 4 utilizes numerical simulation to account for the influence of stellar evolutionary effect and unresolved binaries. Section 5 discusses the impact of different types of SFH on PDMF and the universality of the IMF across the Galactic disk. Finally, we give a summary in Section 6.

## 2. Data

We selected a total of 132,624 early-type stars from M. Xiang et al. (2022), with mass estimated by Q. D. Li et al. (2025). Subsequently, we excluded 3113 sources exhibiting missing Gaia photometry and retained 129,511 early-type stars. The observed sample spans a mass range of 1.3–48 $M_\odot$, a metallicity ([M/H]) range of −0.5–0.5 dex, and includes stars with effective temperatures ($T_{\rm eff}$) hotter than 7000 K. All the stars are located in the main-sequence phase, and we have cross-matched them with Gaia EDR3 (Gaia Collaboration et al. 2021) to obtain Gaia photometry and geometric distance (C. A. L. Bailer-Jones et al. 2021).

To mitigate the significant selection effects associated with varying stellar masses and absolute magnitudes ($M_{\rm G}$) at different distances ($D$), we selected a region with a roughly uniform sample distribution in the $D - M_{\rm G}$ and $D - $ Mass panels, where $D$ means the geometric distance derived from Gaia parallax by C. A. L. Bailer-Jones et al. (2021) and where $M_{\rm G} = G - A_{\rm G} + 5 - 5 \times \log D$, where $A_{\rm G}$ is the extinction in the $G$ band, calculated by S. Wang & X. Chen (2019) and G. M. Green et al. (2019). Figure 1 shows the number density of sample stars. A sufficient number of stars exist when $D$ is between 800 and 6900 pc, the $M_{\rm G}$ is between −3.6 and 4.2 mag, and the mass is between 1.5 and 7.1 $M_\odot$. Therefore, selecting stars within these ranges helps us initially screen for sample completeness. And the following shows the selection criteria applied to the observed sample:

1. 800 pc $\leqslant D \leqslant$ 6900 pc,
2. −3.6 mag $\leqslant M_{\rm G} \leqslant$ 4.2 mag,
3. 1.5 $M_\odot \leqslant$ Mass $\leqslant$ 7.1 $M_\odot$,
4. −2.5 kpc $\leqslant Z \leqslant$ 2.5 kpc,
5. S/N $\geqslant$ 30, and
6. RUWE < 1.4,

where $Z$, S/N, and RUWE represent the distance from the Galactic plane, signal-to-noise ratios of the LAMOST spectra, and renormalized unit weight error from Gaia. Finally, a total of 73,235 early-type stars were selected as the subsample for the analysis of the stellar PDMF.

## 3. PDMF:Correction for LAMOST Selection Effect

In this paper, we describe the mass function using a single power-law function with a power exponent $\alpha$ and a constant $C$:

$$\frac{dN}{dM} = CM^{-\alpha}, \quad (1)$$

where $N$ represents the star count and $M$ is the stellar mass.

A crucial step in deriving the PDMF from the observed sample is to correct for the selection effect of LAMOST. The survey's design, observing conditions, and data processing can introduce biases that affect the measurement of stellar density. T. Cantat-Gaudin et al. (2023) conducted a comprehensive study on the completeness of the Gaia sample. They found that in the densest regions, the completeness drops to approximately 50% at $G \approx 19$ mag, whereas in sparse regions with a high number of observations, the Gaia catalog remains nearly complete down to $G \approx 20.5$ mag. Furthermore, for stars with $G < 18$ mag (the limiting magnitude of the LAMOST survey), the completeness exceeds 95% (see Figure 3 of T. Cantat-Gaudin et al. 2023). Additionally, Gaia also covers the observational footprint of LAMOST. Therefore, we adopted the Gaia sources as the reference data set to correct the selection effect. C. Liu et al. (2017) developed a Bayesian statistical method specifically for LAMOST to estimate the corrected stellar density for each star. The LAMOST selection effect is corrected using the following

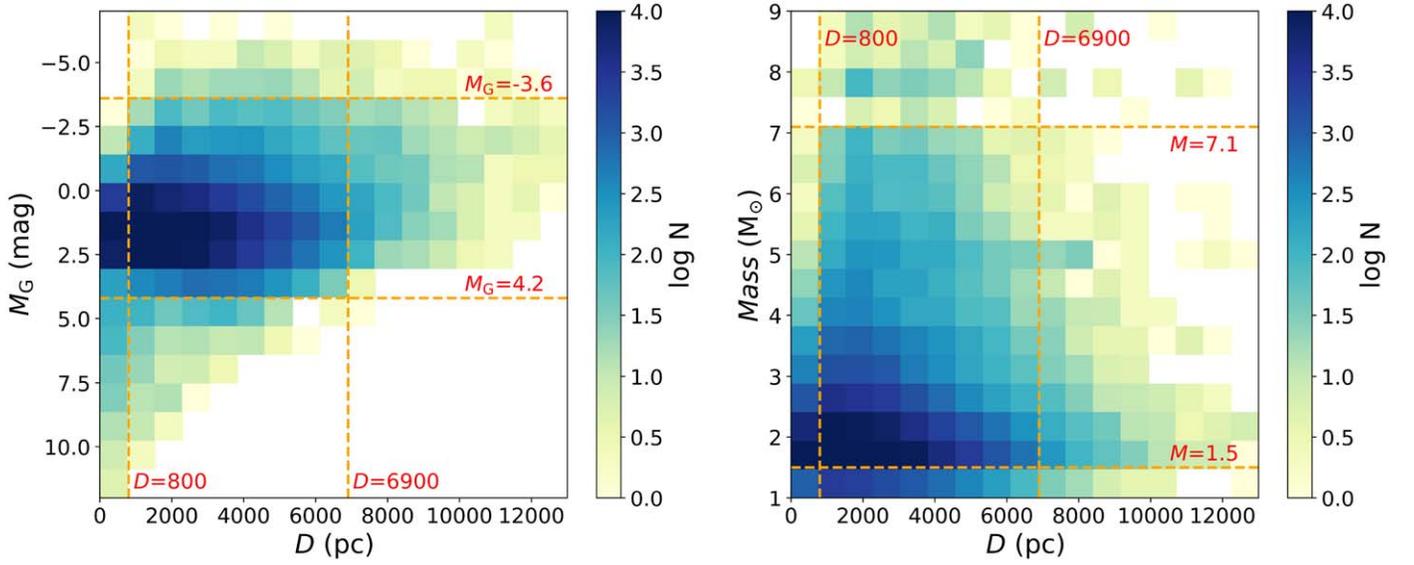

**Figure 1.** Distribution of 129,511 main-sequence stars on the $D - M_G$ (left) and $D - Mass$ (right) panels, with color representing the count in log of each bin. And the orange dashed box highlights the region of the sample with high completeness.

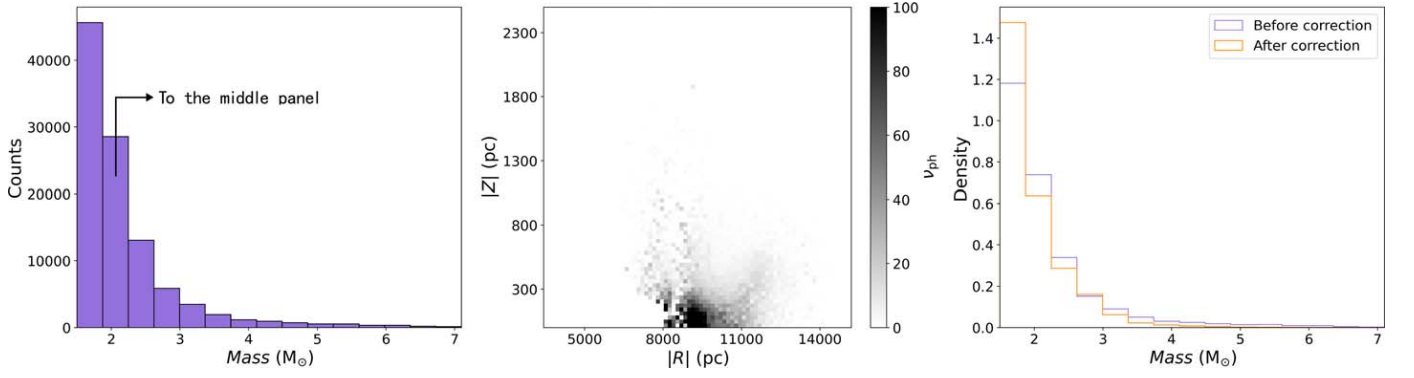

**Figure 2.** Steps to correct the selection effect of LAMOST using the method of C. Liu et al. (2017). The left panel shows the statistical histogram of Mass, the middle panel displays the spatial distribution of stars within the second Mass bin of the left panel, and the right panel compares the normalized results before and after selection-effect correction.

equation:

$$\nu_{\rm ph}(D|c, m, l, b) = \nu_{\rm sp}(D|c, m, l, b) S^{-1}(c, m, l, b). \quad (2)$$

Here, $\nu_{\rm sp}$ is the stellar density from LAMOST data, and $\nu_{\rm ph}$ is the stellar density from Gaia data (ground truth). $c$ and $m$ represent stellar color and magnitude, respectively. $l$ and $b$ are the Galactic longitude and latitude. The selection function $S$ represents the ratio obtained by dividing the number of targets (LAMOST) by the number of photometrically complete samples (Gaia), for each specific magnitude and color:

$$S(c, m, l, b) = \frac{\int_0^\infty \nu_{\rm sp}(D|c, m, l, b) \Omega D^2 \, dD}{\int_0^\infty \nu_{\rm ph}(D|c, m, l, b) \Omega D^2 \, dD}, \quad (3)$$

where $\Omega$ is the solid angle associated with the line of sight (for details, see C. Liu et al. 2017).

Figure 2 illustrates the specific steps involved in correcting for selection effect. First, the subsample is divided into 15 bins based on Mass (left panel). Using the second Mass bin of the left panel as an example, the $\nu_{\rm ph}$ of each star within this Mass bin is calculated using Equation (2). In the middle panel of Figure 2, the distribution of $\nu_{\rm ph}$ in the $|R| - |Z|$ diagram is shown, where $|R|$ is the absolute radial distance and $|Z|$ is the absolute vertical distance from the galactic plane, while the color bar represents the value of $\nu_{\rm ph}$. For each $|R| - |Z|$ bin, the median value of $\nu_{\rm ph}$ is calculated. Then, the sum of the median values for $\nu_{\rm ph}$ in each $|R| - |Z|$ bin gives the total star count for the second Mass bin in the left panel.

Next, we applied the above steps to each Mass bin in the left panel. Finally, the right panel of Figure 2 compares the results before and after the correction for the selection effect. It can be seen that the density of $\sim 1 \, M_\odot$ stars has been increased. This discrepancy can be attributed to the observing strategy or sky conditions, which may have led to the exclusion of faint stars, resulting in a sample that lacks some faint stars before correction.

We then used the maximum likelihood estimation to fit Equation (1) to the observed distribution to obtain the power-law exponent of the PDMF. As shown in Figure 3, the principle of maximum likelihood estimation dictates that the power-law exponent of PDMF ($\alpha_{\rm pre}$) corresponding to the peak of the likelihood represents the estimated value for the observed sample. This analysis yielded an $\alpha_{\rm pre}$ of 5.60 for the observed sample.

To investigate the impact of Mass prediction uncertainty on PDMF measurement, we conducted the following test: We performed 300 Monte Carlo simulations on the initial sample of 129,511 early-type stars from Section 2, using their Mass prediction uncertainties. The selection criteria (Section 2) and the selection-effect correction for the LAMOST sample were applied to the Monte Carlo samples. For each simulation, about 70,000 stars in a mass range from 1.5 to 7.1 $M_\odot$ were used to measure the PDMF. Finally, the effect of Mass prediction errors on PDMF measurement was evaluated. After Monte Carlo sampling, the PDMF we obtained is $a_{\rm pre} = 5.59 \pm 0.02$.

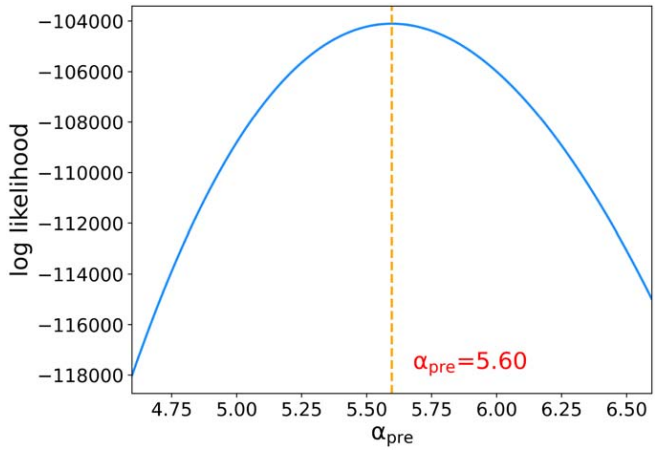

**Figure 3.** Result of fitting a power-law function using the maximum likelihood estimation to the sample after selection-effect correction. The Mass range of the sample is between 1.5 and 7.1 $M_\odot$. The x-axis represents the power-law exponent of the PDMF, and the y-axis represents the logarithm of the likelihood. The peak of the likelihood is highlighted by the orange dashed line, corresponding to an $\alpha_{\rm pre}$ of 5.60.

## 4. Correction of Evolutionary Effect and Unresolved Binary Effect

### 4.1. The Parameters of the Star Formation Simulation Code

With the PDMF of the observed sample, it is necessary to apply a correction for the evolutionary effect and unresolved binary effect. Because our sample is composed of field stars formed through continuous star formation, some of the massive stars formed earlier have evolved off the main sequence. If we do not consider the evolutionary effect, the measured field-star IMF will be very steep. Additionally, the possibility of unresolved binaries within the sample cannot be ruled out. To assess the impact of the evolutionary effect and unresolved binary effect on the mass function, we employ a numerical simulation. We simulated star formation based on the SFH of the Galaxy and calculated stellar evolution using Binary-Star Evolution[5] (BSE; J. R. Hurley et al. 2002). We also considered the effect of unresolved binaries on Mass predictions and star counts. The BSE provides a comprehensive framework for simulating the evolution of binary systems. It incorporates models of stellar structure, mass transfer, and various physical processes to predict the evolution of binary characteristics over time. This enables the study of diverse binary systems and their associated phenomena. In addition, BSE can also be used to calculate the evolution of single stars.

In the Galactic disk, star formation occurs in small bursts along the spiral arms, with most stars forming in stellar systems. Both the cluster mass function and the stellar mass function should be considered in this context (P. Kroupa &

[5] https://astronomy.swin.edu.au/~jhurley/

C. Weidner 2003). Furthermore, the stellar dynamics in the Galactic disk complicate the understanding of the field-star population, as different local volumes contain field stars formed in various locations (I. Minchev et al. 2018). In this work, for simplicity, we assumed that star formation has occurred uniformly across the galactic disk or that stars formed at different epochs and locations are well mixed, as in many previous works (K.-P. Schröder & B. E. J. Pagel 2003; B. G. Elmegreen & J. Scalo 2006; M. A. Czekaj et al. 2014).

Initially, all the stars generated in our simulation are in binary systems. However, binaries with very long orbital periods (generally considered to be over 100 yr) are treated as single stars during evolution. Therefore, our simulation effectively generates both single stars and binary stars. The primary stars are generated continuously from 2.7 Gyr ago to the present, encompassing a Mass range from 1 to 15 $M_\odot$, following a specific IMF (the power-law index ranges from 2.4 to 2.9, with a step of 0.1). The minimum Mass of the subsample is 1.5 $M_\odot$, which corresponds to a maximum lifetime that will not exceed 2.7 Gyr. And the low-mass limit for the primary stars is set to 1 $M_\odot$ to ensure that a single power law can be used for star formation. The metallicity (Z) of the stars is set to the solar value of 0.014 (K. Lodders et al. 2009). In this process, we assume star formation rate (SFR), orbital separation (a), mass ratio (q), and generated distance ($D_{\rm gen}$) of the binary stars, respectively.

(1) *SFR.* Due to the uncertainty of SFR, we tested three SFRs in our code. The SFR from Z. Li et al. (2020) is

$$\text{SFR}(t) = 10.6 \exp\left[\frac{-(t-4)}{9}\right] + 0.125\,(t-4)\,M_\odot\,{\rm yr}^{-1}. \quad (4)$$

The SFR from M. Moe & O. De Marco (2006 is

$$\text{SFR}(t) = f_1 \exp\left[\frac{-(t-3)}{16.6}\right] M_\odot\,{\rm yr}^{-1}. \quad (5)$$

And the SFR from M. A. Czekaj et al. (2014) is

$$\text{SFR}(t) = f_2 \exp(-0.12\,t)\,M_\odot\,{\rm yr}^{-1}, \quad (6)$$

where $f_1 = 8.832$ and $f_2 = 14.762$ are normalization factors to ensure the same total number of stars are formed as that in Equation (4) over the 2.7 Gyr period.

(2) *The orbital separation a.* $a$ is set from the distribution presented in Z. Han et al. (2007):

$$a{\rm n}(a) = \begin{cases} 0.07\,(a/10)^{1.2}, & a \leqslant 10\,R_\odot \\ 0.07, & 10\,R_\odot < a \leqslant 5.75 \times 10^6\,R_\odot \end{cases}. \quad (7)$$

The unit of $a$ is the solar radius ($R_\odot$). This $a$ distribution implies that ~50% of systems are binaries with orbital periods less than 100 yr.

(3) *The mass ratio q.* Two distributions of $q$ are considered. One is a uniform distribution between 0.1 and 1:

$${\rm n}(q) = 1, \quad 0.1 \leqslant q \leqslant 1. \quad (8)$$

Setting the lower limit of $q$ to 0.1 ensures that the Masses of secondary stars are no less than 0.1 $M_\odot$. We also adopted the $q$ distribution for early-type stars presented in M. Moe & R. Di Stefano (2017). In their work, the distribution of $q$ follows a power-law function, and the slope of the power-law function depends on the value of $a$. For each $a$, the slope of the power-law function changes with different ranges of $q$. In Figure 4, the distributions of $q$ for various $a$ values used in this paper are shown.

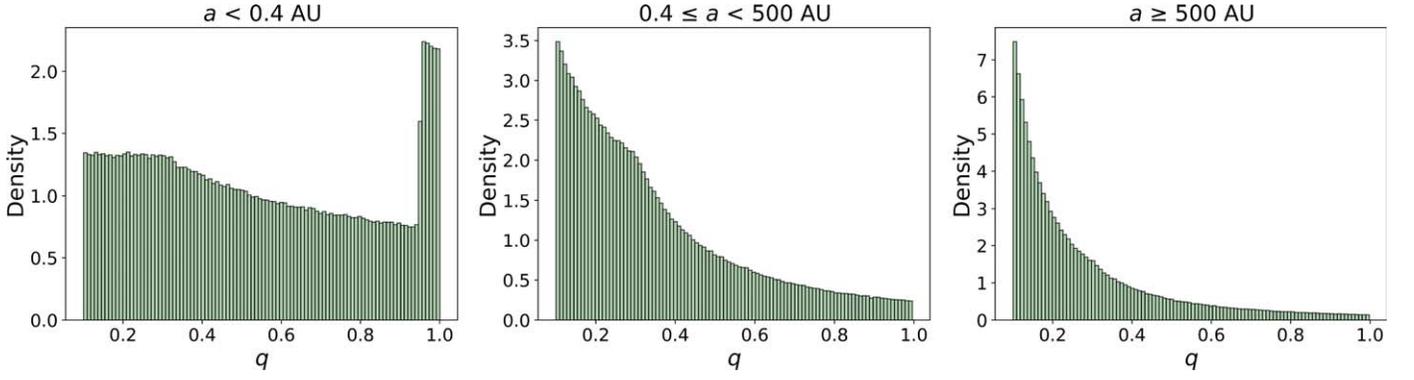

**Figure 4.** $q$ distribution for early-type stars presented in M. Moe & R. Di Stefano (2017). This distribution is a piecewise power-law function with a break at $q = 0.3$. Furthermore, the distribution of $q$ varies depends on the value of $a$.

(4) *The generated distance $D_{gen}$.* $D_{gen}$ is assigned to each generated binary system based on the observed sample, as shown in Figure 5. The purple line represents the observed sample, while the orange line represents the $D_{gen}$ distribution for the generated binary systems, generated using kernel density estimation. The binaries with $a/D_{gen} < 3\!''\!.3$ (the diameters of the LAMOST fibers) are classified as unresolved binaries. An examination of the generated sample revealed that approximately 90% are classified as unresolved binaries.

### 4.2. Workflow of the Star Formation Simulation Code

After setting the code parameters, the simulation was run to model continuous star formation. And the BSE was used to calculate the evolution of both single and binary stars. Finally, the stars that remain in the main sequence until today are selected. This generated sample consists of single stars, resolved binaries, and unresolved binaries. The masses of the single stars and resolved binaries are obtained by the BSE. For unresolved binaries, the procedures for determining the Mass values and conducting star counts are as follows: (1) For the selected unresolved main-sequence binaries ($a/D_{gen} < 3\!''\!.3$), we obtain the luminosities of each component star ($L_1$ and $L_2$) from the BSE. (2) The total luminosity of an unresolved binary is the sum of the luminosities of its primary and secondary stars ($L_{unresolved} = L_1 + L_2$). (3) The unresolved masses ($M_{unresolved}$) of the unresolved binaries are then derived from their total luminosities ($L_{unresolved}$) using the mass–luminosity relation (see Table 1). (4) The combination of the unresolved masses of unresolved binaries ($M_{unresolved}$) with the masses of single stars and resolved binaries yields a simulated sample that closely approximates a real-world observational data set. The PDMF can be obtained by performing a power-law fit to this. This process allows us to simulate the unresolved binary effect.

Additionally, we performed a Kolmogorov–Smirnov test on the observed sample (after selection-effect correction) and the generated sample from BSE (after evolutionary effect and unresolved binary effect) and found that the maximum deviation (D) between the two distributions is about 0.09 with a p-value of 0.16. Figure 6 compares the observed sample with the generated sample. It depicts the cumulative distribution of Mass, where the purple and orange lines represent the observed sample and generated sample, respectively. The generated sample is based on the SFH of Z. Li et al. (2020) and a uniform $q$ distribution.

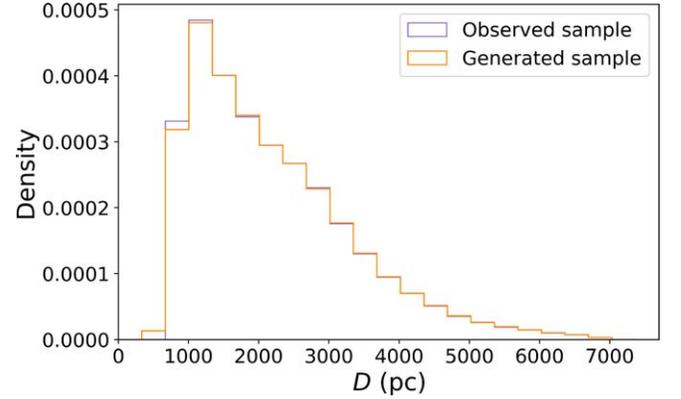

**Figure 5.** Distribution of distance $D$ of the samples. The purple line represents the $D$ distribution of the observed sample, while the orange line represents the $D$ distribution for the generated binary systems, generated using kernel density estimation.

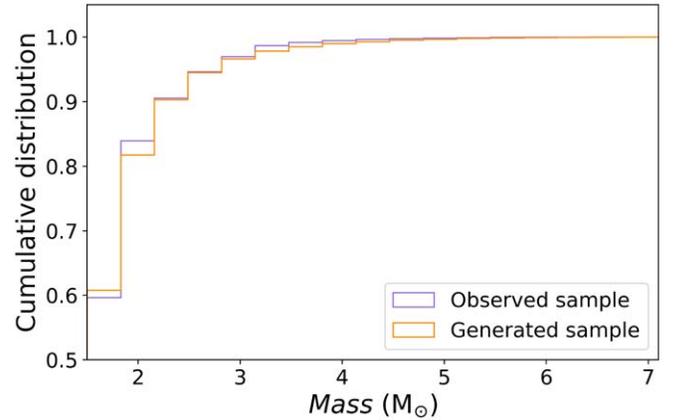

**Figure 6.** Cumulative distribution of Mass for the observed sample (after selection-effect correction) and the generated sample (after evolutionary effect and unresolved binary effect). The purple line represents the observed sample, and the orange line represents the generated sample.

**Table 1**
Mass–Luminosity Relation Used for Simulating the Impact of Unresolved Binaries (from Z. Eker et al. 2015)

| Mass Range ($M_\odot$) | Mass–Luminosity Relation |
| --- | --- |
| $0.38 < M \leqslant 1.05$ | $L/L_\odot \approx 0.94 (M/M_\odot)^{4.84}$ |
| $1.05 < M \leqslant 2.40$ | $L/L_\odot \approx 1.00 (M/M_\odot)^{4.33}$ |
| $2.40 < M \leqslant 7.00$ | $L/L_\odot \approx 1.32 (M/M_\odot)^{3.96}$ |
| $7.00 < M \leqslant 32.00$ | $L/L_\odot \approx 17.26 (M/M_\odot)^{2.73}$ |

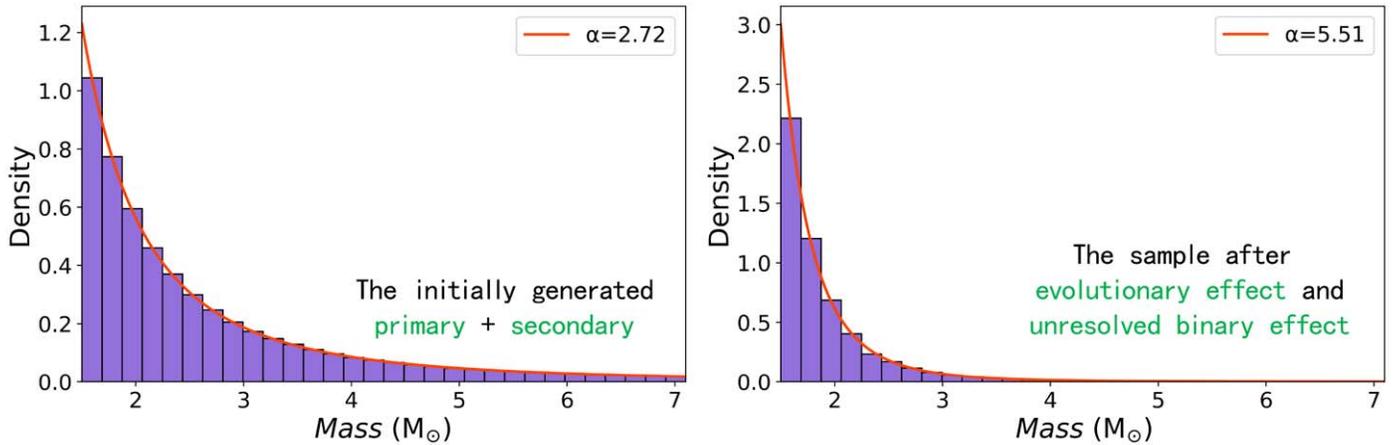

**Figure 7.** Comparison of the initial simulated distribution generated using an IMF of 2.7 (left panel) and the final simulated distribution after accounting for stellar evolution and unresolved binaries (right panel). The reason the slope is not 2.7 in the initial simulated distribution is that generating secondary stars according to the $q$ distribution slightly alters the sample distribution.

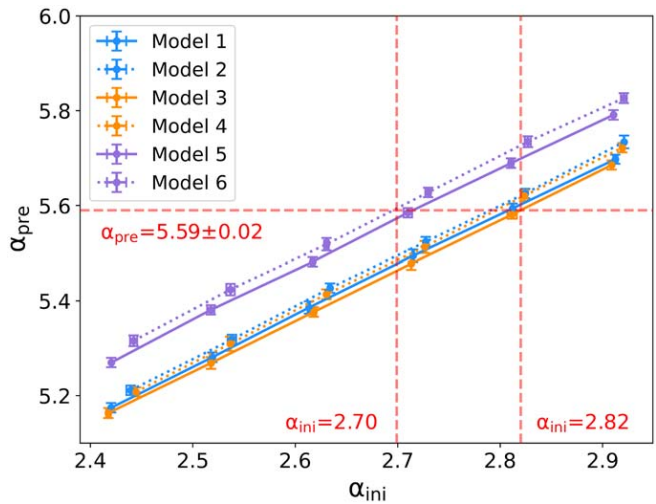

**Figure 8.** Results after correcting for evolutionary effect and unresolved binary effect. The $x$-axis shows the power-law exponent of the IMF, and the $y$-axis shows the power-law exponent of the PDMF. The red dashed lines represent the PDMF and corrected IMFs. Since there are six models, each producing a corrected IMF, only the two extreme values (two red dashed vertical lines) are shown for clarity. The lower and upper limits are 2.70 and 2.82, respectively. Different colored lines represent different SFHs. The dotted lines represent the use of the $q$ distribution from M. Moe & R. Di Stefano (2017), while the solid lines use a uniform $q$ distribution. See Table 2 for details. The analysis only focuses on stars with Mass between 1.5 and 7.1 $M_\odot$.

Finally, by inputting an IMF, we can obtain a corresponding PDMF, establishing a relationship between the IMF and PDMF. This allows us to simultaneously correct for both evolutionary effect and unresolved binary effect. For example, in Figure 7, we generated a simulated sample using an IMF of 2.7 for primary stars, the SFH from Z. Li et al. (2020), and a uniform $q$ distribution. Then we performed a power-law fit to the sample after accounting for stellar evolution and unresolved binary effects. The power-law fit was restricted to stars with Mass between 1.5 and 7.1 $M_\odot$. It is noteworthy that generating secondary stars using the $q$ distribution will slightly alter the power-law form of the simulated sample. Therefore, the values 2.72 and 5.51 exhibit a correspondence. In Figure 8, we present the PDMF of the generated samples obtained under different SFHs and different $q$ distributions. The combinations of SFH and $q$ distribution are listed in Table 2. In Figure 8, the lines shown by the same color represent the same SFH, while different line styles represent different $q$ distributions. The difference in the PDMF between different $q$ distributions under the same SFH and IMF is relatively small, with a $\Delta a_{\text{pre}}$ of $\sim 0.03$. Overall, the uncertainty in SFH has a greater impact on the results. Finally, after accounting for the uncertainties in $q$ and SFH, our final IMF measurement is $\alpha_{\text{ini}} = 2.70 \sim 2.82$.

I. N. Reid et al. (2002) measured the IMF using 558 main-sequence stars, reporting a dispersion of 0.3. K.-P. Schröder & B. E. J. Pagel (2003) derived the IMF from several thousand single stars, with an uncertainty of 0.15. Additionally, D. R. Weisz et al. (2015) measured the IMF for 85 clusters, finding a dispersion of 0.1. In this study, we determined the IMF using 70,000 early-type field stars, achieving a dispersion of 0.12, which is consistent with the typical uncertainties in IMF measurements.

## 5. Discussion

### 5.1. Impact of Different Types of SFH on PDMF

The previous section demonstrates that different SFHs affect the IMF measurements. This uncertainty in SFH primarily manifests as differences in its form, leading to different PDMFs in our star formation simulations. To further investigate this effect, we performed an additional test. Three hypothetical SFHs, with varying types, are used for comparison with the SFH of Z. Li et al. (2020; Equation (4)). Given the uncertainty in the SFH, we referred to M. Cerviño et al. (2016), A. M. Hopkins & J. F. Beacom (2006), and M. Moe & O. De Marco (2006) and subsequently considered three different types of SFHs. They are constant form, linearly decreasing form, and exponentially decreasing form, respectively (see Table 3). The coefficients of these SFH models were adjusted to constrain them to produce the same total stellar mass on the Galactic disk over the past 2.7 Gyr and to better distinguish their steepness. The left panel of Figure 9 shows these four SFHs with different forms. The blue, red, purple, and green lines represent the SFH from Z. Li et al. (2020), constant form, linearly decreasing form, and exponentially decreasing form, respectively. An $x$-axis value of 0 means the present time.

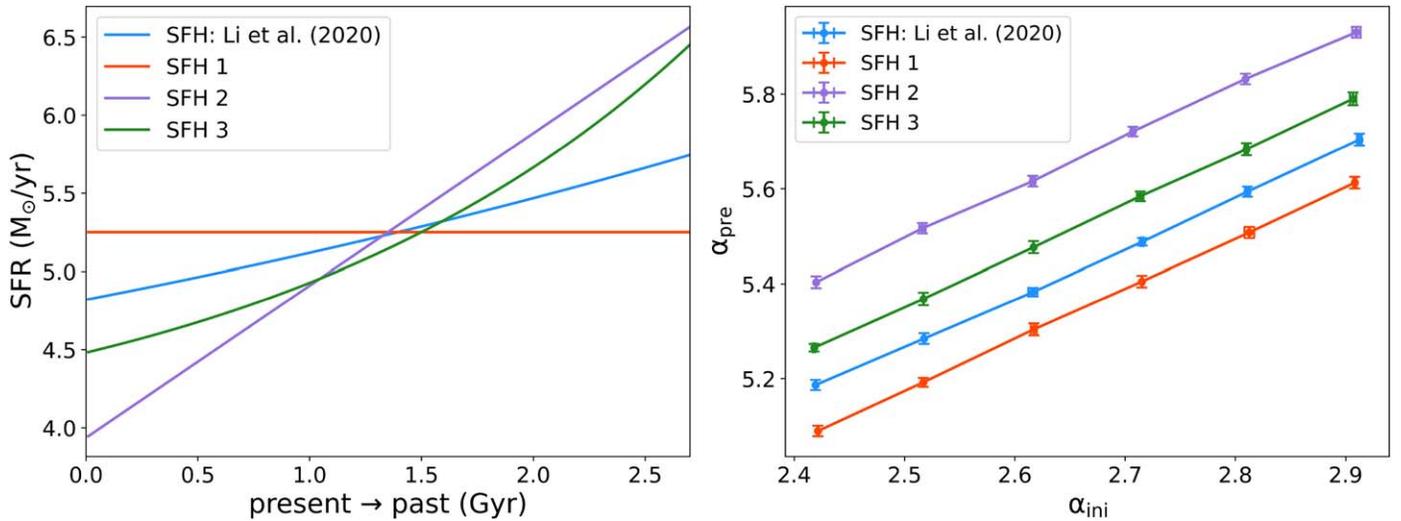

**Figure 9.** Impact of SFH steepness on the PDMF. The left panel shows the SFHs with different steepness levels, normalized to the same total star formation. The blue line is from Equation (4) (Z. Li et al. 2020), the red line represents the constant form, the purple line represents the linearly decreasing form, and the green line represents the exponentially decreasing form. See Table 3 for details. The x-axis means time, with a value of 0 representing the present. The right panel displays the resulting PDMFs after incorporating stellar evolution and unresolved binary effects. The x-axis means the power-law exponent of the IMF, and the y-axis means the power-law exponent of the PDMF.

**Table 2**
Combinations of SFH and $q$ Distribution Used to Obtain the PDMF of the Generated Samples

| Model | Star Formation History | $q$ Distribution |
|---|---|---|
| Model 1 | Equation (4) (Z. Li et al. 2020) | Equation (8) (Uniform) |
| Model 2 | Equation (4) (Z. Li et al. 2020) | Figure 4 (M. Moe & R. Di Stefano 2017) |
| Model 3 | Equation (5) (M. Moe & O. De Marco 2006) | Equation (8) (Uniform) |
| Model 4 | Equation (5) (M. Moe & O. De Marco 2006) | Figure 4 (M. Moe & R. Di Stefano 2017) |
| Model 5 | Equation (6) (M. A. Czekaj et al. 2014) | Equation (8) (Uniform) |
| Model 6 | Equation (6) (M. A. Czekaj et al. 2014) | Figure 4 (M. Moe & R. Di Stefano 2017) |

**Note.** Figure 8 shows the comparison of PDMFs for these combinations.

**Table 3**
Forms of the Three Hypothetical SFHs Used in Testing the Impact of SFH Steepness on PDMF

| Star Formation History | Form | Reference |
|---|---|---|
| SFH 1 | Constant | M. Cerviño et al. (2016) |
| SFH 2 | Linearly decreasing | A. M. Hopkins & J. F. Beacom (2006) |
| SFH 3 | Exponentially decreasing | M. Moe & O. De Marco (2006) |

We then used our star formation simulation code to calculate the results while holding all other parameters the same ($q$ following a uniform distribution). The result of this test is shown in the right panel of Figure 9. With a fixed IMF, a steeper SFH tends to produce PDMF with a larger power-law exponent. B. G. Elmegreen & J. Scalo (2006) investigated the effect of SFH on IMF measurements, and their conclusion is consistent with ours. The number of low-mass stars in all the different SFH models naturally accumulates. However, different models have different degrees of accumulation. Of a population of stars born at some point in the past, only the stars with lower masses remain on the main-sequence and contribute to the statistics in present time. The influence of these stars with lower masses on the statistics depends on their relative abundance in the statistical sample. When the SFH is declining, a steeper decline indicates fewer stars formed recently. This means that the fraction of the stars with lower masses from a point in the past is higher in the overall sample. The result is an increase in the number of stars with lower masses, and the power-law exponent of the PDMF becomes larger.

### 5.2. Universality of the IMF across the Galactic Disk

The universality of the IMF is a subject of ongoing debate in stellar astrophysics. While earlier studies assumed the IMF to be universal, recent evidence suggests that it may vary depending on both metallicity and stellar age (J. Li et al. 2023). These variations challenge the assumption of a universal IMF and highlight the importance of considering environmental factors when studying the mass distribution of stars. To investigate whether the IMF derived from the subsample of the Galactic disk is representative of the entire disk, we applied our analysis to three distinct volume-limited samples within the solar neighborhood: 300, 500, and 800 pc. Each analysis follows the complete procedure described in this study, including corrections for selection effect, evolutionary effect, and unresolved binary effect. This approach allows us to assess the universality of the IMF across different spatial scales. The results are presented in Table 4, showing that the IMF values derived at spatial scales of 300, 500, and 800 pc are $\alpha_{\rm ini} = 2.75 \sim 2.88$, $\alpha_{\rm ini} = 2.68 \sim 2.80$, and $\alpha_{\rm ini} = 2.64 \sim 2.76$, respectively. There appears to be a slight decreasing trend in $\alpha_{\rm ini}$ as the spatial scale increases, which may be related to variations in metallicity. J. Li et al. (2023) found that $\alpha_{\rm ini}$ tends to decrease with decreasing metallicity. Stars farther from the Galactic plane in the vertical direction generally have lower metallicities, which could explain the observed decline in $\alpha_{\rm ini}$ with increasing spatial scale. However, due to limitations in sample size and measurement uncertainties, the results in Table 4 only suggest a potential trend of decreasing $\alpha_{\rm ini}$ with metallicity, and the statistical significance is not particularly high.

**Table 4**
IMFs Derived at Different Spatial Scales

| Spatial Scale (pc) | Sample Size of Stars | IMF |
| --- | --- | --- |
| 300 | 454 | $\alpha_{\rm ini} = 2.75 \sim 2.88$ |
| 500 | 3046 | $\alpha_{\rm ini} = 2.68 \sim 2.80$ |
| 800 | 10,773 | $\alpha_{\rm ini} = 2.64 \sim 2.76$ |

Additionally, several studies have reported results consistent with ours. I. N. Reid et al. (2002) investigated the IMF within 25 pc of the solar neighborhood and found that the high-mass end exhibits $\alpha = 2.5 \sim 2.8$. Similarly, P. Kroupa et al. (2013) reported $\alpha = 2.7 \pm 0.4$ for the Galactic disk for stars with masses greater than $1\,M_\odot$. Given their uncertainties, these results are statistically consistent, further supporting the universality of the IMF across the Galactic disk. However, P. Massey (2002) found that the field-star IMF in the Small and Large Magellanic Clouds exhibits a steeper slope at the high-mass end, with $\alpha \approx 5$. Later, J. B. Lamb et al. (2013) revisited the IMF of OB-type field stars in the Small Magellanic Cloud and reported $\alpha \approx 3.3$. Although the discrepancy between these results may be attributed to sample completeness or measurement uncertainties, both values are significantly steeper than those observed in the Galactic disk. This suggests that the IMF may not be universal across different environments. The field-star IMF reflects the integrated outcome of star formation across different temporal and spatial scales. Therefore, its precise measurement not only enhances our understanding of the fundamental laws of star formation but also provides critical constraints for galaxy evolution models. Despite ongoing research on the field-star IMF, many unresolved questions remain. Future studies could focus on exploring the influence of different star-forming environments on the IMF, such as high SFR environments and low-metallicity environments, which would offer new insights into the variations of the IMF.

## 6. Conclusion

In this paper, we measured the high-mass end of the Galactic-field IMF using ∼70,000 early-type stars from the LAMOST survey. After correcting for the LAMOST selection effect, we obtained the slope of the PDMF to be $\alpha_{\rm pre} = 5.59 \pm 0.02$. We then developed a star formation simulation code to correct for stellar evolution and unresolved binary effects. After considering the uncertainties of $q$ distribution and SFH, we derived the slope of the IMF to be $\alpha_{\rm ini} = 2.70 \sim 2.82$. Furthermore, we investigated the impact of different types of SFH on the PDMF and found that steeper SFH tends to produce PDMF with a larger power-law exponent. Finally, we tested samples across different spatial scales and found that $\alpha_{\rm ini}$ may exhibit a decreasing trend as the spatial scale increases, which could be attributed to variations in metallicity.

This study employed a large sample and statistical methods to investigate the high-mass end of the IMF. We considered the evolution of single and binary stars. In addition, we simultaneously corrected for both evolutionary and unresolved binary effects. These factors improve the reliability of our results, and our work will continue to extend in the future.

### Acknowledgments

This work is supported by the National Natural Science Foundation of China (grant Nos. 12288102, 12125303, 12090040/3), the National Key R&D Program of China (grant Nos. 2021YFA1600401, 2021YFA1600403, 2021YFA1600400), the Natural Science Foundation of Yunnan Province (Nos. 202201BC070003, 202001AW070007), the International Centre of Supernovae, Yunnan Key Laboratory (No. 202302AN360001), the Yunnan Revitalization Talent Support Program-Science & Technology Champion Project (No. 202305AB350003), the China Manned Space Project (No. CMS-CSST-2021-A10), the NSFC (grant No. 12303106), and the Postdoctoral Fellowship Program of CPSF (No. GZC20232976).

### ORCID iDs

Jianping Xiong https://orcid.org/0000-0003-4829-6245
Zhenwei Li https://orcid.org/0000-0002-1421-4427
Dan Qiu https://orcid.org/0000-0002-8280-4808
Chao Liu https://orcid.org/0000-0002-1802-6917
Zhanwen Han https://orcid.org/0000-0001-9204-7778
Xuefei Chen https://orcid.org/0000-0001-5284-8001

### References

Bailer-Jones, C. A. L., Rybizki, J., Fouesneau, M., et al. 2021, AJ, 161, 147
Bastian, N., Covey, K. R., & Meyer, M. R. 2010, ARA&A, 48, 339
Basu, S., & Rana, N. C. 1992, ApJ, 393, 373
Bochanski, J. J., Hawley, S. L., Covey, K. R., et al. 2010, AJ, 139, 2679
Cantat-Gaudin, T., Fouesneau, M., Rix, H.-W., et al. 2023, A&A, 669, A55
Cerviño, M., Bongiovanni, A., & Hidalgo, S. 2016, A&A, 589, A108
Cui, X.-Q., Zhao, Y.-H., Chu, Y.-Q., et al. 2012, RAA, 12, 1197
Czekaj, M. A., Robin, A. C., Figueras, F., et al. 2014, A&A, 564, A102
De Marco, O., & Izzard, R. G. 2017, PASA, 34, e001
Deng, L.-C., Newberg, H. J., Liu, C., et al. 2012, RAA, 12, 735
Eker, Z., Soydugan, F., Soydugan, E., et al. 2015, AJ, 149, 131
Elmegreen, B. G., & Scalo, J. 2006, ApJ, 636, 149
Figer, D. F., Kim, S. S., Morris, M., et al. 1999, ApJ, 525, 750
Gaia Collaboration, Brown, A. G. A., Vallenari, A., et al. 2021, A&A, 649, A1
Graff, D. S., & Freese, K. 1996, ApJL, 467, L65
Green, G. M., Schlafly, E., Zucker, C., et al. 2019, ApJ, 887, 93
Guo, Y., Li, J., Xiong, J., et al. 2022a, RAA, 22, 025009
Guo, Y., Liu, C., Wang, L., et al. 2022b, A&A, 667, A44
Guo, Y., Zhang, B., Liu, C., et al. 2021, ApJS, 257, 54
Han, Z., Podsiadlowski, P., & Lynas-Gray, A. E. 2007, MNRAS, 380, 1098
Hopkins, A. M. 2018, PASA, 35, e039
Hopkins, A. M., & Beacom, J. F. 2006, ApJ, 651, 142
Hosek, M. W., Lu, J. R., Anderson, J., et al. 2019, ApJ, 870, 44
Hurley, J. R., Tout, C. A., & Pols, O. R. 2002, MNRAS, 329, 897
Jeffries, R. D. 2012, EAS Publications Series, 57, 45
Kroupa, P., Tout, C. A., & Gilmore, G. 1993, MNRAS, 262, 545
Kroupa, P., & Weidner, C. 2003, ApJ, 598, 1076

Kroupa, P., Weidner, C., Pflamm-Altenburg, J., et al. 2013, in Planets, Stars and Stellar Systems, ed. T. D. Oswalt & G. Gilmore, 5 (Dordrecht: Springer), 115
Lamb, J. B., Oey, M. S., Graus, A. S., et al. 2013, ApJ, 763, 101
Li, J., Liu, C., Zhang, Z.-Y., et al. 2023, Natur, 613, 460
Li, Q. D., Xiong, J.P., Li, J., et al. 2025, ApJS, 276, 19
Li, Z., Chen, X., Chen, H.-L., et al. 2020, ApJ, 893, 2
Lim, B., Chun, M.-Y., Sung, H., et al. 2013, AJ, 145, 46
Liu, C., Xu, Y., Wan, J.-C., et al. 2017, RAA, 17, 096
Liu, X.-W., Yuan, H.-B., Huo, Z.-Y., et al. 2014, in IAU Symp. 298, Setting the Scene for Gaia and LAMOST (Cambridge: Cambridge Univ. Press), 310
Lodders, K., Palme, H., & Gail, H.-P. 2009, in Landolt-Börnstein - Group VI Astronomy and Astrophysics: Solar System, ed. J. E. Trümper (Heidelberg: Springer), 712
Luhman, K. L. 2012, ARA&A, 50, 65
Massey, P. 2002, ApJS, 141, 81
Miller, G. E., & Scalo, J. M. 1979, ApJS, 41, 513
Minchev, I., Anders, F., Recio-Blanco, A., et al. 2018, MNRAS, 481, 1645
Moe, M., & De Marco, O. 2006, ApJ, 650, 916
Moe, M., & Di Stefano, R. 2017, ApJS, 230, 15
Reid, I. N., & Gizis, J. E. 1997, AJ, 113, 2246
Reid, I. N., Gizis, J. E., & Hawley, S. L. 2002, AJ, 124, 2721
Sagar, R., & Richtler, T. 1991, A&A, 250, 324
Salpeter, E. E. 1955, ApJ, 121, 161
Sana, H., de Mink, S. E., de Koter, A., et al. 2012, Sci, 337, 444
Scalo, J. M. 1986, in IAU Symp. 116, Luminous Stars and Associations in Galaxies (Dordrecht: D. Reidel), 451
Schröder, K.-P., & Pagel, B. E. J. 2003, MNRAS, 343, 1231
Sung, H., & Bessell, M. S. 2004, AJ, 127, 1014
Wang, S., & Chen, X. 2019, ApJ, 877, 116
Weidner, C., Kroupa, P., & Maschberger, T. 2009, MNRAS, 393, 663
Weisz, D. R., Johnson, L. C., Foreman-Mackey, D., et al. 2015, ApJ, 806, 198
Xiang, M., Rix, H.-W., Ting, Y.-S., et al. 2022, A&A, 662, A66